# Evaluation of Digital Forensic Process Models with Respect to Digital Forensics as a Service

Xiaoyu Du, Nhien-An Le-Khac, Mark Scanlon
School of Computer Science,
University College Dublin,
Belfield, Dublin 4, Ireland.

*xiaoyu.du@ucdconnect.ie, {an.lekhac, mark.scanlon}@ucd.ie*

**Abstract:** Digital forensic science is very much still in its infancy, but is becoming increasingly invaluable to investigators. A popular area for research is seeking a standard methodology to make the digital forensic process accurate, robust, and efficient. The first digital forensic process model proposed contains four steps: Acquisition, Identification, Evaluation and Admission. Since then, numerous process models have been proposed to explain the steps of identifying, acquiring, analysing, storage, and reporting on the evidence obtained from various digital devices. In recent years, an increasing number of more sophisticated process models have been proposed. These models attempt to speed up the entire investigative process or solve various of problems commonly encountered in the forensic investigation. In the last decade, cloud computing has emerged as a disruptive technological concept, and most leading enterprises such as IBM, Amazon, Google, and Microsoft have set up their own cloud-based services. In the field of digital forensic investigation, moving to a cloud-based evidence processing model would be extremely beneficial and preliminary attempts have been made in its implementation. Moving towards a Digital Forensics as a Service model would not only expedite the investigative process, but can also result in significant cost savings – freeing up digital forensic experts and law enforcement personnel to progress their caseload. This paper aims to evaluate the applicability of existing digital forensic process models and analyse how each of these might apply to a cloud-based evidence processing paradigm.

1.    **Introduction**

The field of digital forensics has become commonplace due to the increasing prevalence of technology since the late 20$^{th}$ century, and the inevitable relevance of this technology in the conducting of criminal activity. In traditional forensics, the evidence is generally something tangible that could identify the criminal, such as hair, blood or fingerprints. In contrast, digital forensics deals with files and data in digital form extracted from digital devices. Digital forensics is a widely-used term, referring to the identification, acquisition and analysis of digital evidence originating from much more than just computers, such as smartphones, tablets, Internet of Things Devices, or data stored in the cloud.

In the not-so-distant past, most cases involving digital forensic investigation involved criminals using computers, networks or other IT infrastructure as a tool for conducting their crimes. At that time, the set of devices requiring analysis usually consisted of a single computer and the cases involving digital investigation were infrequent. Society has become increasingly reliant on a variety of digital devices, as a result, there is a massively increased need for expert digital forensic analysis across a variety of cases, and a multitude of devices requiring analysis per case has become commonplace. The increasing number of cases involving digital investigation; the number of digital devices requiring analysis is also increasing; the storage volume of each device is growing; the diversity of digital devices and the various form of storage formats, file systems, e.g., Internet-of-Things devices, wearables, cloud storage, etc., introduces additional complexity to the digital forensic process. All these factors ultimately lead to the mounting digital forensic backlog commonly encountered in law enforcement (Lillis et al. 2016).

A standardised framework to guide the process of digital forensics is vital to expedite the process of digital forensic investigation and to address issues such as the increasingly volume of data (Reith et al. 2002; Kohn et al. 2013). Several process models have been defined and refined over time. Each iteration attempted to integrate new technologies and methods over the previous model. The research on process models in recent years, is more concerned with employing new methods and tools into the existing models to improve the efficiency of processing or dealing with the new problem in investigation.

It seems a natural progression for digital forensic processing to move to a cloud environment. Digital Forensics as a Service (DFaaS) is still very much in its infancy, but is already showing significant promise (van Baar et al. 2014). The advantages of migrating to a DFaaS processing model include:
- **Always Up-to-date Software Resources** - A full suite of forensic tools is available in the server support the investigation.
- **Pooled Hardware Resources** - This facilitates increased computing power and storage space.
- **Resources Management** - Only need manage a single system rather than several independent forensic machines.
- **Flexible Location and Time** - Rather than only having the ability to work in the forensic laboratory, investigators will be able to conduct their work remotely from anywhere.

## 1.1 Contribution of this Work

This paper discusses current digital forensic processing models and evaluates their appropriateness and readiness of their applicability to a cloud-based processing model. The contribution of this work can be summarised as follows:
- Discussion of the evolution of digital forensic process models;
- Analysis of the characteristics of each current process models;
- Review current literature on DFaaS;
- Analysing benefits of the DFaaS to the existing process model.

## 2. Literature Review

## 2.1 Process Models

Even though digital forensics is a relatively new research area, it has already made significant progress. The progress is not only from a technology perspective, such as tools to collect and analysis digital evidence, but also with the improvement of methodology. In digital forensics, a process model is the methodology used to conduct an investigation; a framework with a number of phases to guide an investigation. Generally, process models were proposed on the experience of previous work. Due to the variety of cases, e.g., cyber-attacks conducted by IT specialists, civil cases in a corporation, or criminal cases, different investigators tend to follow different methods in their investigative process, there is no standard workflow in digital forensic investigation.

A standard methodology in digital forensics investigation consists of a definition of the sequence of actions necessary in the investigation. A framework, if it is too simplistic or has fewer phases, might not provide much guidance to the investigation process. A framework with more phases and each phase with sub-steps, with more limitation of its usage scenario may prove more useful. Even though it is almost impossible to design a perfect process model that can deal with any investigation, an ideal framework should be general, which means that it could be applied to as many cases as possible. Furthermore, considering that techniques evolve so fast, a well-defined framework should also with the capability to adopt new techniques in the process of investigation.

Numerous process models have been proposed in the literature to date. Generally, each framework attempts to refine the standard methodology for a specific use case and each of these process models take a broadly similar approach. The earliest research concentrated on defining the process of digital forensic investigation (Kohn et al. 2013). More recently, process model research centres around solving more specific issues - specific use cases or focus on particular steps (evidence collection, preservation or examination, analysis). The triage model (Hitchcock et al. 2016; Rogers et al. 2006) is effective for cases that are time sensitive. By employing digital forensics triage, investigators could discover pertinent evidence and the police could get leads about the criminal sooner instead having to wait for the whole report which could take several months or even years.

## 2.2 Digital Forensics as a Service

Cloud computing has become commonplace in today's world. As one example, cloud storage, such as Google Drive, Dropbox, Apple's iCloud, etc., are widely used by consumers around the world. The development of cloud technology is a double-edged sword from a digital forensic perspective; the wide use of cloud infrastructure and applications brings complexity to conducting digital forensic invesitigations, while leveraging this on-demand, high-speed technology could also make much of the investigative process significantly more efficient. However, based on the current literature in the area, 'Cloud Forensics' is much more popular, i.e., recovering evidence from cloud services and applications. Research on DFaaS is still quite limited in the digital forensic community.

DFaaS is very much still in its infancy. In the last decade, many corporations have finished their processing and data migration from their own servers to the cloud service vendors, such as Amazon or Rackspace. Likewise, in the process of digital forensic investigation, DFaaS could bring several improvements over the existing process.

Lee & Un (2012) outline a number of benefits of making use of cloud computing for forensic investigation. Firstly, remotely connecting with powerful servers instead of each single device not only offer the investigators stronger computing power for evidence examination but also get rid of the location limitation that the analysis could only conducted in the laboratory.

## 3. The Evolution of Digital Forensic Process Models

Several process models have been proposed to date. Current models can be categorised into three main types:
- The first type consists of general models that define the entire process of digital forensic investigation. These models were proposed from 2000 to 2010. Through that time, precisely what should be done and the order to do each step in a digital forensic investigation was still somewhat controversial.
- The second type focus on a particular step in the investigation process or a specific kind of investigative case;
- The third type defined new problems and/or explored new methods or tools to address specific issues.

### 3.1 Early Digital Forensic Process Models

At the turn of the century, it was still the early days of research on digital forensics and digital forensic process models. Initially, one of the most urgent issues in digital forensics was to define a process model to make the entire investigative process consistent and standardised. A number of general digital forensic processing models have been defined. Most of these frameworks define a group of necessary steps in a whole investigation process, and the models were refined over time. The later models improve upon the former ones by including some additional steps or defining sub-steps of the process models - making each step more precisely defined.

The traditional framework had been refined and formed a number of novel frameworks. Some inheritance relation among the existing frameworks listed below:
- *DFRWS model* (Palmer et al. 2001) **=>** *SRDFIM* (Agarwal et al. 2011)
- *DFRWS model* (Palmer et al. 2001) **=>** *An Abstract Digital Forensics Model* (Reith et al. 2002)
- *IDIP (Carrier et al. 2003) & DCSA (Rogers 2006)* **=>** *CFFTPM (Rogers et al. 2006)*
- *Integrated Digital Investigation Process (IDIP) (Carrier & Spafford 2004) => Enhanced Integrated Digital Investigation Process(EIDIP) (Baryamureeba & Tushabe 2004)*
- *Integrated Digital Forensic Process Model* (Kohn et al. 2013) **=>** *DFaaS Process Model* (van Baar et al. 2014)

The focus of these models is to define the phases on typical investigations, the sequence of these phases and the definition of the key concepts of each phase (Palmer et al. 2001; Lee et al. 2001; Reith et al. 2002; Baryamureeba & Tushabe 2004; Beebe & Clark 2005).

Henry Lee proposed a Scientific Crime Scene Investigation (SCSI) model for digital forensic investigation in 2001 (Lee et al. 2001). Ciardhuáin (2004) criticises the SCSI model is not a systematic digital forensic process model as it only focuses on physical crime scene investigation and lack of describing on digital criminal scene investigation. Kohn et al. (2013) explained that the physical crime scene investigation process can be adapted to digital crime scene investigation. The Event-based Digital Forensic Investigation Framework separates the concepts of the physical crime scene and the digital crime scene, collecting digital devices from the physical crime scene and then obtaining digital evidence from the digital devices' storage (Carrier & Spafford 2004). In 2000, Casey defined a digital forensic process model and was refined further in 2004. Casey's model focuses on digital evidence processing and examining. The Enhanced Integrated Digital Investigation Process (EIDIP) model was proposed by Baryamureeba & Tushabe (2004). The EIDIP model is based on IDIP, and introduces a traceback phase to address the problem of having to reconstructing twice in IDIP.

Figure 1 lists out each phase and sub-phase of the aforementioned frameworks:

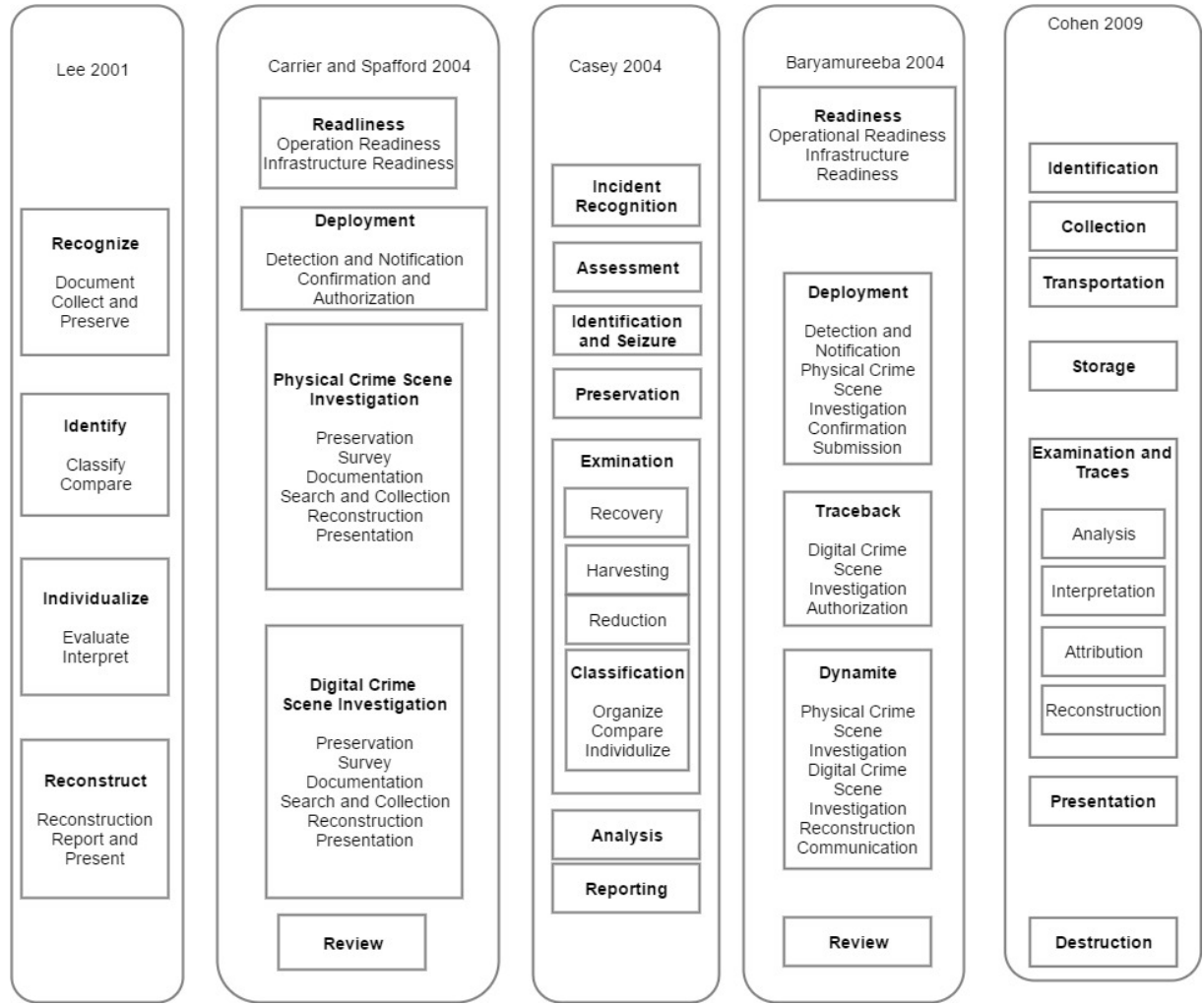

Figure 1: Proposed Digital Forensic Framework in Initial Phase

### 3.2  Refining Digital Forensic Process Models

Merely following a general process model is often not specific enough to handle the broad range of cases typically encountered by law enforcement. The criminal could be an IT specialist and conduct advanced cybercrimes, CCTV cameras' storage may need to be analysed, or data leakage in a corporation, etc. These different situations often require bespoke methodologies.

After the general process procedure was clearly defined, researchers started working on specific issues that are more detailed. For example: 1) refining a process model by make an improvement at a specific step of the investigation; 2) dealing only with a specific category of cases, such as, network forensics, mobile devices forensics, etc.; 3) Triage models (Rogers et al. 2006; Hitchcock et al. 2016) outline specific processes for time sensitive cases, such as child abductions, missing person cases, etc.

The phases and sub-phases of these process models are shown in Figure 2 below:

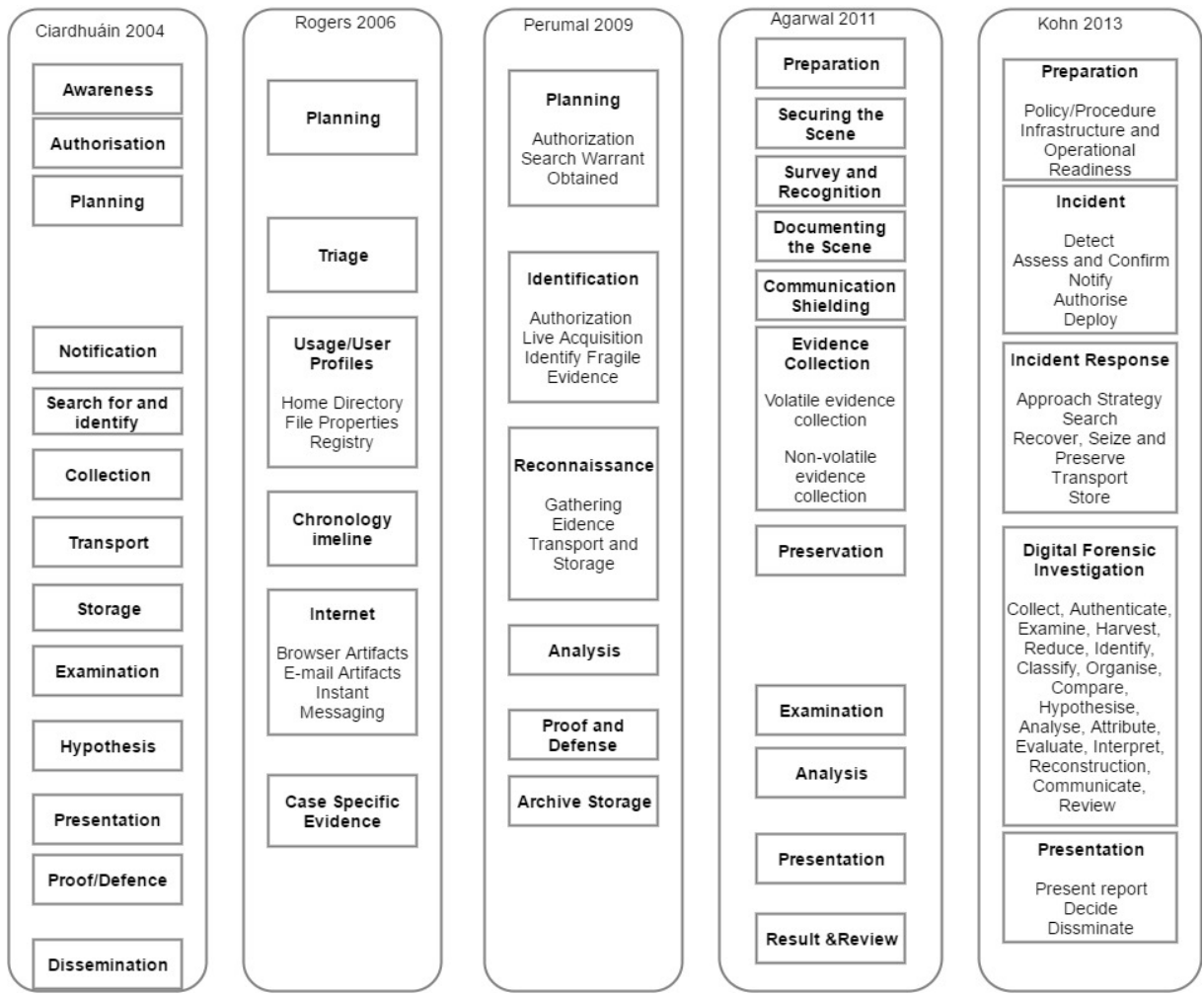

Figure 2: Digital Forensics Frameworks Focusing on a Specific Use Cases

    A. **Extended Model of Cybercrime Investigation** - In 2004, several process models had already been defined. However, each did not include a significant aspect of cybercrime investigation itself. An extended model of cybercrime investigation was proposed by Ciardhuáin (2004). This model follows a waterfall fashion and the necessary activities are conducted in sequence. This model allows iteration in some part of the investigation, for example, the iterative process of "examination - hypothesis - presentation - proof/defence".

    B. **Digital Forensic Triage Process Model** - In some special cases, such as kidnaps and hostage rescue, acquiring clues from digital devices immediately is crucial, or some other cases such as robbery, crucial information is required as soon as possible to increase the likelihood of catching the criminal before they have escaped to another country. Often traditional models are insufficient for this use case - potentially taking weeks or years to get results. Tiered models are designed to expedite situations like this. Considering traditional models are designed to guide the entire investigation, a triage process model was proposed to deal with time sensitive cases (Rogers et al. 2006). This model focuses on the crucial first few hours of an investigation.

    C. **Digital Forensic Model Based on Malaysian Investigation Process -** This model is notable in that it is focused on data acquisition process, including more detailed handling on live data acquisition and static data acquisition in cybercrime investigation (Perumal 2009).

    D. **The Systematic Digital Forensics Investigation Model** - This model is focus on computer fraud and cybercrimes, which is helpful in evidence dynamics and reconstruction (Agarwal et al. 2011).

E.  **Integrated Digital Forensic Process Model -** This model is the most recent proposed process model which including a relative generally digital forensic investigation (Kohn et al. 2013).

## 3.3 Recent Research on Digital Forensic Process Models

Some new and popular technologies result in new problems hindering digital forensics investigations. Cloud computing makes evidence collection more difficult; Internet-of-Things adds a variety of new device and storage forms; more digital devices connected into the Internet result in an ever-increasing volume of data. In recent years, research on process models is more focused on integrating other technologies, such as data mining, to support the original models, or propose novel process models to solve the issues caused by these new technologies.

Some recent models, as outlined in Figure 3, include:
- An integrated conceptual digital forensic framework for cloud computing (Martini & Choo 2012).
- Data reduction and data mining framework (Quick & Choo 2014).
- Internet of Things (IoT) Based Digital Forensic Model (Perumal et al. n.d.).

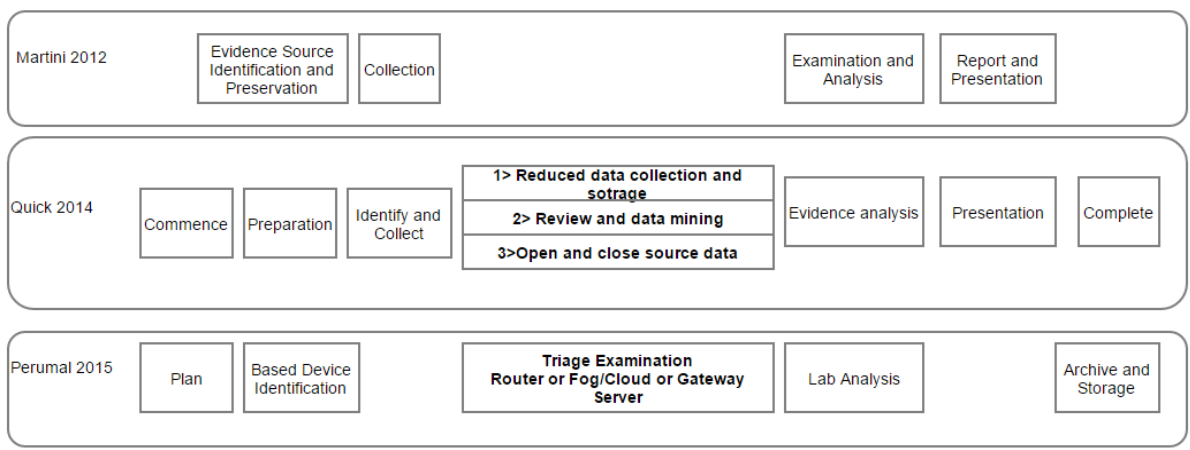

Figure 3: Recent Digital Forensic Models for Handling Modern Advancements

A.  **An Integrated Conceptual Digital Forensic Framework for Cloud Computing -** As the prevalence of cloud computing services increases, collecting digital evidence from a remote server, which often is stored in another jurisdiction, has become necessary. In recent years, researchers in digital forensics have been trying to address the issues encountered in Cloud Forensics. An integrated conceptual digital forensic framework was proposed by Martini and Choo (2012) based on two widely used basic models: (McKemmish 1999) and (Kent et al. 2006).

The difficulties encountered conducting a forensic investigation of a cloud service can be identified in each stage of a typical case. Firstly, the determination that cloud forensics is necessary might only be possible after acquiring cached information or stored login credentials from a physical digital device, such as a laptop or smartphone. It is as if the investigator opens one door (physical digital evidence devices) and gets a key of the other (cloud evidence). If the first key was not discovered (e.g., lost through mishandling of volatile data), there is no possibility to get the second key. As the result, the investigator would never retrieve any evidence behind the second door. Secondly, in the collection of cloud evidence, the problems often found include: 1) no possibility to physically seizing all the servers in a cloud computing environment; 2) the server could be in another jurisdiction; 3) the collection of metadata might not be possible; etc.

B.  **Data Reduction and Data Mining Framework** - Considering the new challenges encountered in digital forensic investigation, Quick and Choo (2014) list seven requirements of forensic analysis: *faster collection, reduced storage, timely review, intelligence, research, knowledge management, archive and retrieval*. One challenge in digital forensics is the ever-increasing volume of data, which has impeded investigations from a number of standpoints including evidence collection, data preservation and analysis. The growth of digital evidence has been ongoing for many years and is safely predicted to increase further into the future.

The core idea of this framework is to acquire a subset of the data by utilising data reduction and conduct intelligence analysis through data mining. Obviously, the subset prioritises files which are the most crucial and important for investigation. This subset is much smaller than the entirety of the evidential data, and as a result, any operations investigators conduct on it would be significantly faster. This subset of data could bring number of significant benefits for investigation:

- Triage devices and media;
- Faster indexing;
- Provide potential to utilise data mining or intelligence analysis;
- Cross-case analysis;
- Enable research of historical case data and intelligence analysis.

C. **Internet of Things Based Digital Forensic Model** - The growing prevalence of Internet-of-Things (IoT) brings with it new problems for digital forensics. As a new challenge in this area, the volume of digital devices needing to be collected, analysed, examined and preserved, as well as the variety of storage formats make analysis more arduous. A more sophisticated forensic model, which aims to address the specific issues relating to IoT based investigation, is that proposed by Perumal et al. (2015). This model defines a standard operating procedure for investigation of IoT devices.

D. **Field Processing Model** – One of the more recent models proposed surrounds a digital forensic field processing model (Hitchcock et al. 2016). This model is focused on training non-digital evidence to specialists conducting the early stage of investigation on scene. The front-line investigators analyse the pertinent information first and a more detailed examination and analysis will be subsequently conducted in the laboratory. This research on one hand solves the problem of the shortage of digital forensic specialists in law enforcement, and on the other hand helps relieve the digital forensic backlog. Coupling DFaaS with this field triage processing model could result in significant benefits. Namely, the traditional laboratory-based examination could be conducted on scene through a laptop connected with the cloud system. This would afford the investigator the use of a powerful computing resource in the field.

4. **Digital Forensics as a Service**

Even though, cloud computing has become prevalent across many industries, there is limited literature on its use and advantages from a DFaaS perspective (Lee and Un 2012; van Baar et al. 2014; Wen et al. 2013). In this section, the current research on DFaaS will be discussed.

The first utilisation is the computing power provided by distributed computing, which can better handle the increasing magnitude of data. Lee & Un (2012) shows the efficiency of cloud system working on indexed search. Wen et al. (2013) outline an implementation of cloud based system to combat the magnitude of data encountered by digital forensics by leveraging parallel computing. This work highlights the applicability of cloud computing in digital forensics and the improvement that DFaaS could make.

One use case of DFaaS is to offer indexed search as a service (Lee & Un 2012). Concerning the large volume of data needing to be analysed, distributed computing systems could do the same work in parallel. Such cloud server can offer highly intensive computing process and large quantity of storage to deal with the slow processing on big data volume. In their paper, Lee and Un outline a case study that indexed search as a service.

In 2013, Wen et al. designed a cloud based framework, which *deals with large volume of forensic data, sharing interoperable forensic software, and providing tools for forensic investigators to create and customise forensics data processing workflows.* After a series of tests, the experimental results show that *the proposed workflow management solution can save up to 87% of analysis time in the tested scenarios*. In this framework, the main purpose of making use of cloud systems to deal with the large volume of evidence data through distributed parallelisation.

In 2014, van Baar et al. outlined an implementation of DFaaS in the Netherlands Forensic Institute. It focused on a comparison between the DFaaS framework with the traditional models and list the problems in traditional methodology while outlining how their DFaaS implementation has addressed some of these issues. This work proves the viability and impact cloud-based digital forensic solutions can have on the entire process.

## 4.1 Benefits and Advantages of DFaaS

There is little doubt that DFaaS can improve the efficiency of the investigative process. The growing volume of data results in an increased time needed for each step of a typical digital forensic investigation. Leveraging cloud computing with its significant computing resources would be one obvious solution to this issue. A centralised data storage server could expedite the process of evidence collection and analysis (Scanlon 2016). In addition, a cloud-based digital forensics environment could enable case detectives to directly connect and perform preliminary analysis themselves in a controlled environment without waiting for expert analysis. In this triage model, DFaaS facilitates the investigators preserve and analyse digital evidence on scene by connecting to the server remotely. The management of forensics environment would still ultimately be handled by digital forensic specialists.

A broadly applicable framework that can deal with numerous existing situations encountered in digital forensics, while being extensible to handle new technologies has always been desirable. DFaaS enables this to be possible. DFaaS not only benefits from the processing power cloud computing provides, but can also influence future development of digital forensic science – opening up new possibilities for collaborative investigation. The evidence from cases could be stored into the cloud-based system, making more intelligent forensic processing possible (Quick and Choo 2014). New tools and techniques developed in the field of structured and unstructured data science could also be easily integrated to further expedite the process.

## 5. Conclusion

A rigorous methodology and a standardised procedure of investigation process is vital for conducting forensic investigations. The pursuit of a perfect model for digital forensic science will likely never cease. In this paper, the evolution of digital forensic process models was discussed and these models were classified into three types. The first type defines a general process for the entire investigation process. The second type refines and enhances the previous models by improving compatibility with more situations. The third type makes use of new methods, techniques and/or tools in the investigative process to deal with new problems encountered in modern investigations. Overall, future refinements of the digital forensic process will likely focus on usage scenarios, improving the efficiency of the investigative process, and incorporating new technologies and techniques into the models for the purposes of ensuring an always adaptable methodology.

## 5.1 Future Work

Society is increasingly moving their day-to-day life to the digital world. The huge volume of data has created several challenges for digital forensics. By using theories and tools from data science to address these challenges in digital forensics is a valuable research direction in digital forensics. Considering the significant influence which DFaaS could make in digital forensics, future work will focus on building an extensible processing model focusing on the cloud-based handling of digital evidence.